\documentstyle[epsf]{mn}

\newif\ifAMStwofonts

\ifoldfss
  \ifCUPmtlplainloaded \else
    \NewTextAlphabet{textbfit} {cmbxti10} {}
    \NewTextAlphabet{textbfss} {cmssbx10} {}
    \NewMathAlphabet{mathbfit} {cmbxti10} {} 
    \NewMathAlphabet{mathbfss} {cmssbx10} {} 
  \fi
  \ifAMStwofonts
    \ifCUPmtlplainloaded \else
      \NewSymbolFont{upmath} {eurm10}
      \NewSymbolFont{AMSa} {msam10}
      \NewMathSymbol{\upi}     {0}{upmath}{19}
      \NewMathSymbol{\umu}     {0}{upmath}{16}
      \NewMathSymbol{\upartial}{0}{upmath}{40}
      \NewMathSymbol{\leqslant}{3}{AMSa}{36}
      \NewMathSymbol{\geqslant}{3}{AMSa}{3E}

    \fi
  \fi
\fi 

\ifnfssone
  \newmathalphabet{\mathit}
  \addtoversion{normal}{\mathit}{cmr}{m}{it}
  \addtoversion{bold}{\mathit}{cmr}{bx}{it}
  \newmathalphabet{\mathbfit} 
  \addtoversion{normal}{\mathbfit}{cmr}{bx}{it}
  \addtoversion{bold}{\mathbfit}{cmr}{bx}{it}
  \newmathalphabet{\mathbfss} 
  \addtoversion{normal}{\mathbfss}{cmss}{bx}{n}
  \addtoversion{bold}{\mathbfss}{cmss}{bx}{n}
  \ifAMStwofonts
    \ifCUPmtlplainloaded \else
      \UseAMStwoboldmath
      \makeatletter
      \new@mathgroup\upmath@group
      \define@mathgroup\mv@normal\upmath@group{eur}{m}{n}
      \define@mathgroup\mv@bold\upmath@group{eur}{b}{n}
      \edef\UPM{\hexnumber\upmath@group}
      \new@mathgroup\amsa@group
      \define@mathgroup\mv@normal\amsa@group{msa}{m}{n}
      \define@mathgroup\mv@bold\amsa@group{msa}{m}{n}
      \edef\AMSa{\hexnumber\amsa@group}
      \makeatother
      \mathchardef\upi="0\UPM19
      \mathchardef\umu="0\UPM16
      \mathchardef\upartial="0\UPM40
      \mathchardef\leqslant="3\AMSa36
      \mathchardef\geqslant="3\AMSa3E
    \fi
  \fi
\fi 

\ifnfsstwo
  \DeclareMathAlphabet{\mathbfit}{OT1}{cmr}{bx}{it}
  \SetMathAlphabet\mathbfit{bold}{OT1}{cmr}{bx}{it}
  \DeclareMathAlphabet{\mathbfss}{OT1}{cmss}{bx}{n}
  \SetMathAlphabet\mathbfss{bold}{OT1}{cmss}{bx}{n}
  \ifAMStwofonts
    \ifCUPmtlplainloaded \else
      \DeclareSymbolFont{UPM}{U}{eur}{m}{n}
      \SetSymbolFont{UPM}{bold}{U}{eur}{b}{n}
      \DeclareSymbolFont{AMSa}{U}{msa}{m}{n}
      \DeclareMathSymbol{\upi}{0}{UPM}{"19}
      \DeclareMathSymbol{\umu}{0}{UPM}{"16}
      \DeclareMathSymbol{\upartial}{0}{UPM}{"40}
      \DeclareMathSymbol{\leqslant}{3}{AMSa}{"36}
      \DeclareMathSymbol{\geqslant}{3}{AMSa}{"3E}
    \fi
  \fi
\fi 

\ifCUPmtlplainloaded \else
  \ifAMStwofonts \else 
    \def\upi{\pi}
    \def\umu{\mu}
    \def\upartial{\partial}
  \fi
\fi

\title[ELAIS paper VI: Discovery of a new hyperluminous infrared galaxy]
{The European Large Area \iso Survey VI - Discovery of a new hyperluminous infrared galaxy}
\author[T. Morel, \etal]
{\parbox{179mm}{\begin{flushleft}
\vspace{-0.5cm}
{\LARGE T. Morel,$^1$}
\thanks{Present address: Inter-University Centre for Astronomy and Astrophysics (IUCAA), Post Bag 4, Ganeshkhind, Pune, 411 007, India; e-mail: morel@iucaa.ernet.in}
{\LARGE A. Efstathiou,$^1$}
{\LARGE S. Serjeant,$^{1,2}$}
{\LARGE I. M\'arquez,$^3$}
{\LARGE J. Masegosa,$^3$}
{\LARGE P. H\'eraudeau,$^4$}
{\LARGE C. Surace,$^1$}
{\LARGE A. Verma,$^{1,5}$}
{\LARGE S. Oliver,$^{1,6}$}
{\LARGE M. Rowan-Robinson,$^1$}
{\LARGE I. Georgantopoulos,$^7$}
{\LARGE D. Farrah,$^1$}
{\LARGE D. M. Alexander,$^{8,9}$}
{\LARGE I. P\'erez-Fournon,$^{10}$}
{\LARGE C. J. Willott,$^{11}$}
{\LARGE F. Cabrera-Guerra,$^{10}$}
{\LARGE E. A. Gonzalez-Solares,$^{10}$}
{\LARGE A. Cabrera-Lavers,$^{10}$}
{\LARGE J. I. Gonzalez-Serrano,$^{10,12}$}
{\LARGE P. Ciliegi,$^{13}$}
{\LARGE F. Pozzi,$^{13}$}
{\LARGE I. Matute,$^{14}$}
{\LARGE and H. Flores$^{15}$}
\end{flushleft}
}\vspace*{0.200cm}\\  
\parbox{159mm}{
$^1$ Astrophysics Group, Imperial College of Science, Technology and Medicine, Blackett Laboratory, Prince Consort Road, London, SW7 2BZ, UK\\
$^2$ Unit for Space Sciences and Astrophysics, School of Physical Sciences, University of Kent, Canterbury, Kent, CT2 7NR, UK\\   
$^3$ Instituto de Astrof\'{\i}sica de Andaluc\'{\i}a, CSIC, Apartado 3004, E-18080, Granada, Spain\\
$^4$ Max-Planck-Institut f\"{u}r Astronomie, K\"{o}nigstuhl 17, D-69117 Heidelberg, Germany\\
$^5$ Max-Planck-Institut f\"{u}r Extraterrestrische Physik, Postfach 1312, D-85741 Garching, Germany\\
$^6$ Astronomy Centre, University of Sussex, Falmer, Brighton, BN1 9QJ, UK\\
$^7$ Astronomical Institute, National Observatory of Athens, Lofos Koufou, Palaia Penteli, Athens, GR-15236, Greece\\
$^8$ SISSA, International School for Advanced Studies, Via Beirut 2-4, 34014, Trieste, Italy\\
$^9$ Pennsylvania State University, Astronomy and Astrophysics, Davey
Laboratory, University Park, PA 16802, USA\\
$^{10}$ Instituto de Astrof\'{\i}sica de Canarias, Via Lactea, 38200 La Laguna, Tenerife, Canary Islands, Spain\\
$^{11}$ Astrophysics, Department of Physics, Keble Road, Oxford, OX1 3RH, UK\\
$^{12}$ Instituto de F\'{\i}sica de Cantabria, (CSIC --- Univ. de Cantabria), Facultad de Ciencias, Univ. de Cantabria, 39005, Santander, Spain\\
$^{13}$ Osservatorio Astronomico di Bologna, via Ranzani 1, 40127, Bologna, Italy\\
$^{14}$ Dipartimento di Fisica, Universita degli Studi ``Roma TRE'',
Via della Vasca Navale 84, I-00146, Roma, Italy\\
$^{15}$ CEA/SACLAY, 91191 Gif sur Yvette cedex, France}}

\date{Accepted ???.
      Received ???;
      in original form ??? }
\pagerange{\pageref{firstpage}--\pageref{lastpage}}
\pubyear{2001}

\begin{document}
\def\is{{\it IRAS} }
\def\ir{{\it IRAS}}
\def\m{$\mu$m}
\def\ms{$\mu$m }
\def\etal{et al. }
\def\es{ELAISP90\_J164010+410502 }
\def\e{ELAISP90\_J164010+410502}
\def\eas{ELAIS J1640+41 }
\def\ea{ELAIS J1640+41}
\def\iso{{\it ISO} }

\maketitle

\label{firstpage}

\begin{abstract}
\normalsize We report the discovery of the first hyperluminous infrared galaxy (HyLIG) in the course of the European Large Area \iso Survey (ELAIS). This object has been detected by \iso at 6.7, 15, and 90 \m, and is found to be a broad-line, radio-quiet quasar at a redshift: {\it z} = 1.099. From a detailed multi-component model fit of the spectral energy distribution, we derive a total IR luminosity: $L_{\rm IR}$ (1-1000 $\mu$m) $\approx$ 1.0 $\times$ 10$^{13}$ $h_{65}^{-2}$ L$_{\odot}$ ($q_0$ = 0.5), and discuss the possible existence of a starburst contributing to the far-IR output. Observations to date present no evidence for lens magnification. This galaxy is one of the very few HyLIGs with an X-ray  detection. On the basis of its soft X-ray properties, we suggest that this broad-line object may be the face-on analogue of narrow-line, Seyfert-like HyLIGs.
\end{abstract}


\section[]{Introduction}
Early ground-based infrared (IR) observations have revealed the existence of a population of galaxies emitting the bulk of their bolometric energy in this wavelength range (e.g., Rieke \& Low 1972). Further studies of these objects (see Sanders \& Mirabel 1996 for a review and definition of the various sub-types) have revealed their potential importance in the context of galaxy evolution (e.g., Sanders \etal 1988) or in contributing to the cosmic far-IR/sub-mm background (e.g., Hughes \etal 1998). Observational characteristics of this class of galaxies are now relatively well-established. In particular, interacting processes are widely believed to play a pivotal role in triggering the IR excess, not only in the local Universe (Sanders \& Mirabel 1996 and references therein) but also at moderately high redshift (Smail \etal 1998). While a contribution to the IR output can arise {\it both} from a dust-enshrouded active galactic nucleus (AGN) and from a (generally circumnuclear) starburst, there is strong evidence from optical or mid-IR spectroscopy for an increasing fraction of objects with AGN-like characteristics when progressing toward higher luminosities (e.g., Lutz \etal 1998; Veilleux, Sanders, \& Kim 1999). 

Of particular relevance for a better understanding of this important population are the objects at the upper end of the luminosity distribution (the so-called hyperluminous infrared galaxies; hereafter HyLIGs).\footnote{We adopt in the following the definition of Rowan-Robinson (2000): $L$(1-1000 \m) $>$ 10$^{13}$ $h_{65}^{-2}$ L$_{\odot}$ ($q_0$ = 0.5). We use the same cosmology and $H_0$ = 65 km s$^{-1}$ Mpc$^{-1}$ throughout this paper.} 
If their luminosities are starburst-dominated, they are potentially forming the bulk of their stellar population in a single, violent episode. This makes
them an interesting sub-population of galaxies, but work so far
is severely hampered by inhomogenous and/or AGN-biased selection. Only 13 HyLIGs are known at present from far-IR or sub-mm surveys, and a further 12 from cross-correlating far-IR source lists with
AGN catalogues (Rowan-Robinson 2000; hereafter RR). 

In regard of the exceedingly small space density of HyLIGs (van der Werf \etal 1999; RR), only major extragalactic IR surveys offer good prospects to uncover new candidates. Of particular interest in this respect is the European Large Area \iso Survey (ELAIS) which was conducted over $\approx$ 11 square degrees at 15 \ms and 90 \ms (down to $\approx$ 3 mJy and $\approx$ 100 mJy, respectively) and over 6 square degrees at 6.7 \ms (down to $\approx$ 1 mJy). We refer the reader to Oliver \etal (2000; Paper I) for a complete description of this project. Details on the technical aspects of the CAM and PHOT observations can be found in Serjeant \etal (2000; Paper II) and Efstathiou \etal (2000a; Paper III), respectively. Assuming pure luminosity evolution of the form: (1 + {\it z})$^{3.1}$ (Boyle, Shanks, \& Peterson 1988; Saunders \etal 1990), we estimate that about 3 HyLIGs might be detected in the total area covered by ELAIS. This estimate is considerably uncertain, as it is strongly dependent on the spectral energy distribution (SED) and evolutionary model assumed. As we can safely rule out no-evolution models, however, we find that the formal significance of detecting one object is very high ($\approx$ 35$\sigma$). We report in this paper on the discovery of such a candidate HyLIG in the ELAIS N2 region. 

\section{Identification of the new HyLIG}
This new HyLIG was discovered as part of a study aiming at studying the IR colour properties of ELAIS galaxies (Morel \etal 2001). The procedure used to isolate galaxies potentially detected at 6.7, 15, and 90 \ms was to cross-correlate the CAM and PHOT source lists using a 35$''$ association radius. Details on the production of the CAM and PHOT catalogues can be found in Paper II and H\'eraudeau \etal (2001), respectively.

Figure 1 shows the PHOT error circle of the HyLIG (\e; hereafter \ea) overlayed on a deep (down to {\it r}' $\approx$ 24) optical image. The CAM error circles are also shown. The only obvious optical counterpart to the \iso sources is found to be a point-like object at: $\alpha$ = 16$^{\rm h}$40$^{\rm m}$10.16$^{\rm s}$ and $\delta$ = +41$^{\circ}$05$'$22.3$''$ (J2000). This source is, within the uncertainties, spatially coincident with an optically-selected quasar at a quoted redshift of 1.097 ([CCS88] 163831.4+411107; Crampton \etal 1988). As can be seen in Fig.1, a bright optical source (as well as a number of much fainter ones) appear to lie in close vicinity of the quasar. Since no redshifts are available for these objects, it is unclear at this stage whether they are physically associated with the new HyLIG.

\begin{figure}
\epsfxsize=10cm
\epsfysize=10cm
\epsfbox{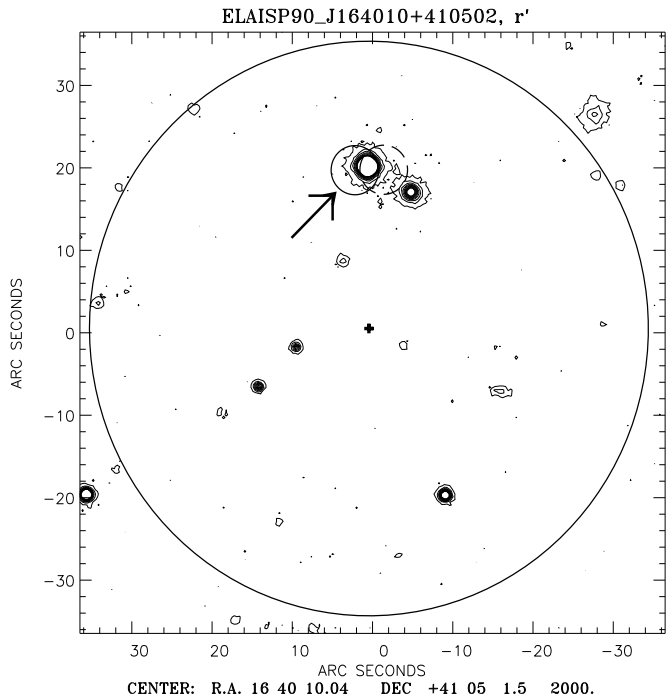}
\vspace*{-0.5cm}
\caption{Optical ({\it r}') contour map of \eas centered on the PHOT detection ({\it cross}). The lowest contour is drawn at 3$\sigma$, and then by steps of 6$\sigma$. The error circle for PHOT (35$''$), as well as the 6$''$ error circles for the detections at 6.7 \ms ({\it solid}) and 15 \ms ({\it dashed}) are also overlayed. The new HyLIG is indicated by an arrow.}
\end{figure} 

We used the quasar
{\it B}-band source counts of Wisotzki \etal (2000) to estimate
the number of random PHOT-quasar associations in our sample at {\it B} = 17.2 mag (see Table 2), or brighter. The expected total number of
random associations in the search radius is only 0.011 for the 285
 PHOT sources. Furthermore, 35\% of the ELAIS N2 area is covered by
the parent sample of the quasar, i.e., only about 10\% of the total ELAIS
area. The total number of random associations in the overlap region is
therefore only $\approx$ 1 $\times$ 10$^{-3}$, giving strong support for the
reality of the PHOT-quasar association.

The quasar does not appear in the \is FSC or FSR catalogues. To test the reliability of our PHOT detection, we made \is ADDSCANs of the source, as well as of 15
galaxies with reliable 90 \ms detections at similar flux densities (60-80 mJy)
from the Paper III catalogue, and 16 control positions at random 20$'$
offsets from the target positions. We found our
PHOT 90 \ms data to be not well reproduced by the ADDSCAN 100 \ms
measurements, although most of the flux densities were
positive. However, the control fields gave values distributed around zero
as expected, from which we derived a $\approx$ 4$\sigma$ ADDSCAN detection of
our PHOT source. This confirms the reality of the source, but we also
conclude that it is too faint to reliably measure the flux density
with ADDSCANs. We only consider in the following upper limits from the FSC.

Optical spectroscopy was obtained at the Nordic Optical Telescope using ALFOSC\footnote{The Nordic Optical Telescope (NOT) is operated on the island of La Palma jointly by Denmark, Finland, Iceland, Norway, and Sweden, in the Spanish Observatorio del Roque de los Muchachos of the  Instituto de Astrof\'{\i}sica de Canarias. ALFOSC is owned by the Instituto de
      Astrof\'{\i}sica de Andaluc\'{\i}a (IAA) and operated at the NOT under agreement
      between IAA and the NBIfAFG of the Astronomical Observatory of Copenhagen.} ({\it R} $\approx$ 700) in non-photometric conditions and seeing of 1.5$''$. We used a 1.2$''$-wide slit oriented EW. The exposure time was 3 $\times$ 900 s. The spectrum was reduced using standard {\tt IRAF}\footnote{{\tt IRAF} is
distributed by the National Optical Astronomy Observatories, operated by
the Association of Universities for Research in Astronomy, Inc., under
cooperative agreement with the National Science Foundation.} routines and is shown in Figure 2. We derive a redshift: {\it z} = 1.099 $\pm$ 0.002. This value is consistent with previous estimates (Crampton, Cowley, \& Hartwick 1989), and is used in the following. Table 1 gives the relative line fluxes, EWs, and FWHMs.

\begin{table}
\caption{Line properties of \ea} 
\begin{tabular}{lccc}
\hline\hline
Line & Relative line flux                                                & $W_{{\rm rest}}$    & FWHM \\
     & & (\AA) & (km s$^{-1}$)\\\hline
C III] $\lambda$1909$^a$ & 8.0 $\pm$ 4.0 & 19.7 $\pm$ 3.0 & 6800 $\pm$ 300\\
C II] $\lambda$2327 & 1.0 & 4.40 $\pm$ 2.0 & 3500 $\pm$ 600\\
Mg II $\lambda$2798 & 3.5 $\pm$ 1.5 & 21.0 $\pm$ 3.0 & 3000 $\pm$ 300\\\hline
\end{tabular}\\
NOTES TO TABLE 1 ---\\
$^a$: Deblended from Al III $\lambda$1857. Might receive a contribution from Si III] $\lambda$1892.
\end{table}

\begin{figure}
\epsfxsize=8.5cm
\epsfysize=8.5cm
\epsfbox{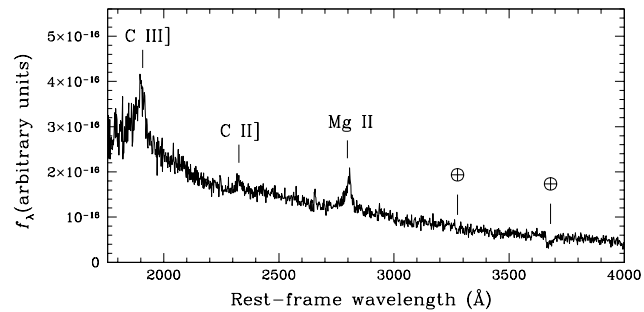}
\vspace*{-4.0cm}
\caption{Rest-frame UV/optical spectrum of \eas (assuming a redshift of 1.099). The main lines are indicated (the $\oplus$ symbols mark terrestrial telluric absorption features). The spectrum has been corrected for both atmospheric and foreground galactic extinction. We assumed $E(B-V)$ = 0.006 mag (Schlegel, Finkbeiner, \& Davis 1998).}
\end{figure} 

\section[]{Model fit of the SED}
Table 2 summarizes the available photometric data for this seldom-studied object. New sub-mm observations (see Farrah \etal 2001) were carried out with the Submillimetre Common-User Bolometer Array (SCUBA) at the James Clerk Maxwell Telescope (JCMT) in photometry mode. Data reduction was performed using the standard
SURF pipeline software. The source is undetected both at 450 and 850 \m, with 3$\sigma$ upper limits of 81.6 and 6.96 mJy, respectively. The region surrounding the HyLIG has unfortunately not been observed as part of the radio survey of the northern ELAIS areas (Ciliegi \etal 1999), but the available radio flux limits imply that this quasar is radio-quiet (Sopp \& Alexander 1991). Following V\'eron-Cetty \& V\'eron (2000), we obtain: $M_{\rm B}$ $\approx$ -- 26.0 mag. 

\begin{table}
\caption{Photometric properties of \ea} 
\begin{tabular}{lll}
\hline\hline
Band & Flux density/Mag. & Source/Ref.\\\hline
0.1-2.4 keV$^a$   & 1.9 $\times$ 10$^{-5}$ mJy       & {\it ROSAT}\\
{\it U}           & 16.362 $\pm$ 0.200             & INT\\
{\it B}$^b$           & 17.17 $\pm$ 0.25        & USNO\\
{\it r}'          & 16.94 $\pm$ 0.05    & INT\\
{\it R}           & 16.686 $\pm$ 0.200      & INT\\
{\it J}           & 16.395 $\pm$ 0.134    & 2MASS\\
{\it H}           & 15.520 $\pm$ 0.139    & 2MASS\\
{\it K$_{\rm S}$}       & 15.164 $\pm$ 0.180    & 2MASS\\
ISOCAM-LW2$^c$      & 4.99   $\pm$ 0.66 mJy & Morel \etal (2001)\\
12 \ms      & $<$200 mJy (3$\sigma$) & \is FSC\\
ISOCAM-LW3$^c$       & 7.75   $\pm$ 0.72 mJy & Morel \etal (2001)\\
25 \ms      & $<$200 mJy (3$\sigma$) & \is FSC\\
60 \ms  & $<$200 mJy (3$\sigma$)     & \is FSC\\
ISOPHOT-C90      & 72   $\pm$ 23 mJy & Morel \etal (2001)\\
100 \ms      & $<$1000 mJy (3$\sigma$) & \is FSC\\
450 \ms      & $<$81.6 mJy (3$\sigma$) & SCUBA\\
850 \ms      & $<$6.96 mJy (3$\sigma$) & SCUBA\\
4.85 GHz & $<$18 mJy (5$\sigma$)                & WENSS \\
1.4 GHz & $<$0.95 mJy (5$\sigma$)                & FIRST\\\hline
\end{tabular}\\
NOTES TO TABLE 2 ---\\
The optical and near-IR magnitudes have been corrected for foreground galactic extinction assuming $E(B-V)$ = 0.006 mag (Schlegel \etal 1998) and the conversion factors of Cardelli, Clayton, \& Mathis (1989).\\
$^a$: Assuming continuum emission with a photon index $\Gamma$ = 2 (see Reeves \& Turner 2000) and a foreground galactic extinction with a column density: $\cal N_{\rm H}$ = 1.1 $\times$ 10$^{20}$ cm$^{-2}$ (Stark \etal 1992).\\ 
$^b$: We do not adopt the value quoted by Crampton \etal (1988): {\it B} = 17.9 $\pm$ 0.2 mag, as it is discrepant both with APM and USNO magnitudes. This might be partly due to quasar variability.\\
$^c$: These values are subject to a possible systematic scaling by up to a factor of about 1.5 (see Paper II).
\end{table}

The SED is displayed in Figure 3. Assuming that the host galaxy is similar to coeval quasars (e.g., McLure \etal 1999), we used the radio galaxy {\it K}-{\it z} relation of Eales \etal (1997) to estimate the host galaxy near-IR magnitude, obtaining: $K$ = 17.0 $\pm$ 0.4 mag. As can be seen in Fig.3, the contribution of the host galaxy to the SED is negligible. In order to estimate the energy budgets to the IR luminosity, we modelled the SED by simultaneously considering models for dust-enshrouded AGNs (Efstathiou \& Rowan-Robinson 1995) and starbursts (Efstathiou, Rowan-Robinson, \& Siebenmorgen 2000b). Although we do not claim to have exhaustively
explored the possible physical parameter space of these models, we found that the SED is most naturally explained by a combination of these two components. Pure AGN models have difficulties in accounting for the far-IR emission. We achieve a satisfactory fit to the SED (see Fig.3) by assuming roughly similar contributions from dust grains heated by the nuclear UV/optical continuum ($L_{\rm IR}^{\rm AGN}$ $\approx$ 5.2 $\times$ 10$^{12}$ $h_{65}^{-2}$ L$_{\odot}$) and by a starburst ($L_{\rm IR}^{\rm SB}$ $\approx$ 5.0 $\times$ 10$^{12}$ $h_{65}^{-2}$ L$_{\odot}$). The difficulties in fitting the region around 3-7 \ms might be related to uncertainties in the ISOCAM absolute calibration (Paper II).

\begin{figure}
\epsfxsize=8cm
\epsfysize=8cm
\epsfbox{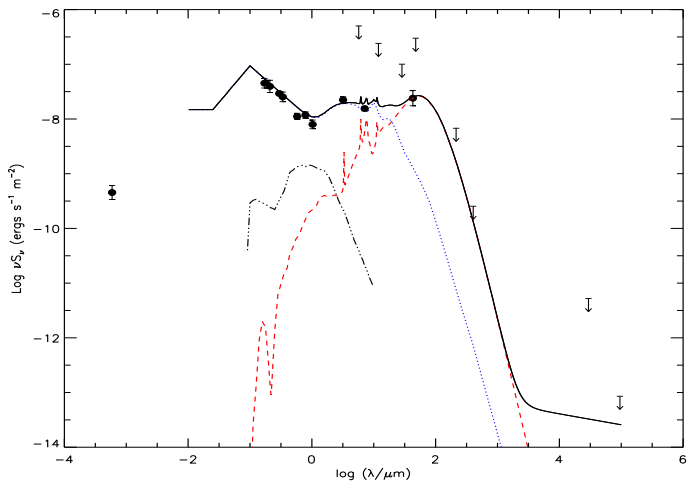}
\vspace*{-1.5cm}
\caption{Rest-frame SED of \eas from X-rays to radio (data from Table 2).
The SED is fitted with a combination of one of the starburst
models of Efstathiou \etal (2000b) ({\it dashed line}; $\tau_V$ = 50, $t$ = 26 Myrs)
and an AGN dusty torus model ({\it dotted line}) from Efstathiou \&
Rowan-Robinson (1995). We assume a half-opening angle for the torus of
45$^{\circ}$ and a face-on inclination. The starburst SED is
extrapolated in the radio using the far-IR/radio correlation (Helou, Soifer, \& Rowan-Robinson 1985). For the host galaxy, the SED of a 4 Gyr old stellar population ({\it dotted-dashed line}; Fioc \& Rocca-Volmerange 1997) has been normalized to {\it K} = 17.0 $\pm$ 0.4 mag (see text).}
\end{figure} 

The {\it total} rest-frame IR luminosity we derive: $L_{\rm IR}$ (1-1000 $\mu$m) $\approx$ 1.02 $\times$ 10$^{13}$ $h_{65}^{-2}$ L$_{\odot}$, is slightly above the threshold for a HyLIG classification in the scheme of RR. The robustness of the derived luminosity has been explored by considering two low-redshift objects in the list of RR with detections in all \is bands (IRAS 07380--2342 and IRAS 18216+6418). Similarly to the approach adopted here, the IR luminosities have been derived by RR from a multi-component model fit of the SED, and are found to agree to within 25\% with the values determined by the method prescribed by Sanders \& Mirabel (1996). We conclude that estimates of the total IR luminosity are fairly insensitive to the choice of the method used, although the relatively low value we obtain for \eas (coupled with the quite large uncertainties in the IR fluxes) indicates that this object can only be formally considered as a candidate HyLIG at this stage. 

Gravitational lensing is known to enhance the luminosity in a number of HyLIGs (e.g., Serjeant \etal 1995). While our data do not allow us to rule out lens magnification at this stage, this possibility can be explored in the future via high-resolution {\it HST} imaging or by seeking for intervening Mg II
absorbers in high-quality, 
moderate-resolution spectroscopic data (e.g., Goodrich \etal 1996).

\section{Discussion}
\subsection{\eas in relation to other HyLIGs}
We compare in Figure 4 the SED of \eas with a sample of HyLIGs with optical evidence for quasar activity. It can be seen that \eas (which is the least luminous in this subset) presents a relative excess in the far-IR. Interestingly, the strongest outlier (BR 1202--0725) is also the most luminous (RR). This trend of hotter colour temperature with increasing luminosity might suggest an increasing contribution from AGN emission at high luminosities (see also Haas \etal 2000). However, such an effect is not apparent in the sample of quasars studied by Polletta \etal (2000). The detection of a large reservoir of molecular gas in BR 1202--0725 (Ohta \etal 1996), but not in PG 1634+706 (Barvainis \etal 1998) also seems to be at variance with this simple picture. 

\begin{figure}
\epsfxsize=8cm
\epsfysize=8cm
\epsfbox{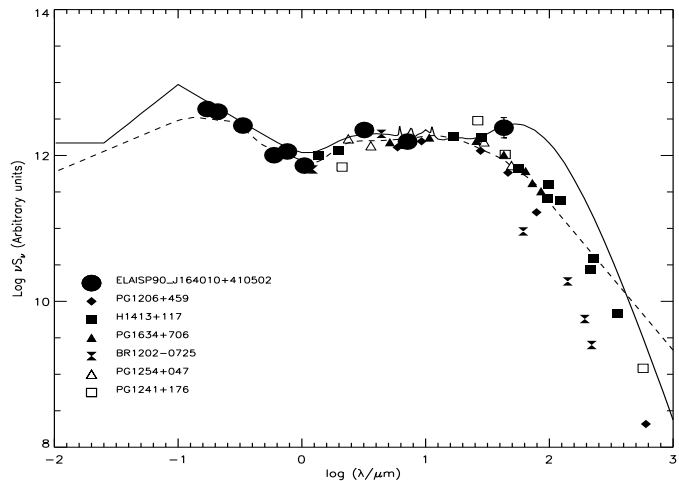}
\vspace*{-1.5cm}
\caption{Comparison of the rest-frame SEDs of a number
of HyLIGs with optical evidence for quasar activity (data from RR and references
therein). The solid and dashed lines show the model fitted to \eas (see Fig.3) and the mean SED of radio-quiet quasars in the UVSX sample (Elvis \etal 1994), respectively. The data points are
normalized to their rest-frame flux at 15 \m. The vertical
scale is arbitrary.}
\end{figure}

We also show in Fig.4 the mean SED for radio-quiet quasars in the UVSX sample (Elvis \etal 1994). It can be seen that \eas displays a significant far-IR excess compared to this population, hence giving credence to the existence of a starburst contributing to the IR output. Indicative of the unsually strong level of IR emission from \eas is the fact that a number of more powerful quasars in the optical regime are {\it not} member of the HyLIG class (McMahon \etal 1999). Assuming a Salpeter initial mass function, we derive from the luminosity of the starburst component a current star-formation rate in the range: 1-7 $\times$ 10$^{3}$ $h_{65}^{-2}$ M$_{\odot}$ yr$^{-1}$, depending upon the stellar low- and high-mass cutoffs assumed (see Thronson \& Telesco 1986). Although such high values have already been claimed in HyLIGs with quasar activity (e.g., McMahon \etal 1999), we caution that the paucity of far-IR detections does not allow us to set stringent constraints on the star-formation rate in \ea. 

Current samples of HyLIGs are severely affected by selection effects. General conclusions regarding the characteristics of this population must thus await the completion of large, unbiased surveys. With this limitation in mind, there is some indication at present for a diversity of power supplies. While AGN-related emission is able to account for the IR output in some HyLIGs (e.g., Granato, Danese, \& Franceschini 1996), others show evidence for a dominant contribution from star formation (e.g., Ivison \etal 1998). The results of our two-component model fit to the SED of \eas supports a composite nature for this object. The existence of a number of HyLIGs whose IR luminosity can be accounted for by a combination of an AGN and a starburst component has been postulated on similar grounds by RR and Verma \etal (2001). The inference of a starburst component mostly contributing to the rest-frame flux longward of about 50 \ms is independently supported in several cases by the large amount of molecular material deduced from radio CO observations (RR).

\subsection{Soft X-ray properties of \ea}
Inspection of the {\it ROSAT} all-sky survey 0.1-2.4 keV photon
maps shows that our object lies within the 90\% error box 
 of an X-ray source detected at the 3.4$\sigma$ confidence level. We obtain a rest-frame (uncorrected for intrinsic absorption) X-ray luminosity in the 0.1-2.4 keV band: $L_{\rm  X}$ $\approx$ 3.7 $\times$ 10$^{44}$ ergs s$^{-1}$, which translates into a $L_{\rm X}$/$L_{{\rm bol}}$ ratio of about 9.4 $\times$ 10$^{-3}$. This ratio supports an AGN interpretation for the soft X-ray emission (Wilman \etal 1998), but is almost two orders of magnitude higher than the upper limits found for HyLIGs with Seyfert 2-like optical spectra (Lawrence \etal 1994; Fabian \etal 1996; Wilman \etal 1998). The spectral index of the UV/optical power-law continuum in \eas ($\alpha$ $\approx$ -- 0.04; $f_{\nu}$ $\propto$ $\nu^{\alpha}$) is typical of UV-selected quasars (e.g., Natali \etal 1998), and suggests little dust obscuration along our line-of-sight to the nucleus. In contrast, we estimate that column densities: ${\cal N}_{\rm H}$ $\ga$ 2 $\times$ 10$^{23}$ cm$^{-2}$ would attenuate the (redshifted) $\approx$ 3-keV emission of \eas to levels comparable to what is observed in Seyfert 2-like HyLIGs.\footnote{See {\tt http://asc.harvard.edu/toolkit/pimms.jsp}} This value compares well with the typical amount of obscuring material intrinsic to these objects (e.g., Wilman \etal 1998). In the context of the unification scheme, the soft X-ray properties of \eas are thus consistent with this galaxy being a face-on analogue of such narrow-line HyLIGs (see Franceschini \etal 2000). This interpretation is independently supported in several cases by the detection of strongly polarized continuum emission in edge-on objects (e.g., Hines \etal 1995).

\section*{Acknowledgments}
We wish to thank K. G. Isaak for carrying out the sub-mm observations, as well as the referee (D. Sanders) and M. Villar-Martin for useful comments. This paper is based on observations with {\it ISO}, an ESA project with
instruments funded by ESA 
member states (especially the PI countries: France, Germany, the
Netherlands and the United 
Kingdom) and with participation of ISAS and NASA. This work was
supported by PPARC (grant 
number GR/K98728) and by the EC TMR Network programme
(FMRX-CT96-0068). This publication makes use of data products from the Two Micron All Sky Survey, which
is a joint project of the University of Massachusetts and the Infrared Processing and Analysis
Center/California Institute of Technology, funded by the National Aeronautics and Space Administration 
 and the National Science Foundation. This research has made use of the NASA/IPAC Extragalactic Database (NED) which is operated by the Jet Propulsion Laboratory, California Institute of Technology, under contract with the National Aeronautics and Space Administration.

\bsp

\label{lastpage}

\end{document}